\documentclass[%
        final,
        notitlepage,
        narroweqnarray,
        inline,
        ]{ieee}
\usepackage{ieeefig}
\usepackage{amsfonts}
\usepackage{epsfig}
\usepackage{graphicx}
\usepackage{stfloats}

\begin{document}

%----------------------------------------------------------------------
% Title Information, Abstract and Keywords
%----------------------------------------------------------------------
\title[DPC for MIMO CRC with Imperfect CSIT]{%
       Dirty Paper Coding for the MIMO Cognitive Radio Channel with Imperfect CSIT}

% format author this way for conference proceedings
\author[C. S. Vaze and M. K. Varanasi]{%
Chinmay S. Vaze and Mahesh K. Varanasi
    \thanks{%
This work was supported in part by NSF Grants CCF-0431170 and
CCF-0728955. The authors are with the Department of Electrical and
Computer Engineering, University of Colorado, Boulder, CO 80309-0425
USA (e-mail: {Chinmay.Vaze, varanasi}@colorado.edu, Ph:
001-303-492-7327).
      }
}

% make the title
\maketitle

% do the abstract
\begin{abstract}
A Dirty Paper Coding (DPC) based transmission scheme for the
Gaussian multiple-input multiple-output (MIMO) cognitive radio
channel (CRC) is studied when there is imperfect and perfect channel
knowledge at the transmitters (CSIT) and the receivers,
respectively. In particular, the problem of optimizing the sum-rate
of the MIMO CRC over the transmit covariance matrices is dealt with.
Such an optimization, under the DPC-based transmission strategy,
needs to be performed jointly with an optimization over the
inflation factor. To this end, first the problem of determination of
inflation factor over the MIMO channel $Y=H_1 X + H_2 S + Z$ with
imperfect CSIT is investigated. For this problem, two iterative
algorithms, which generalize the corresponding algorithms proposed
for the channel $Y=H(X+S)+Z$, are developed. Later, the necessary
conditions for maximizing the sum-rate of the MIMO CRC over the
transmit covariances for a given choice of inflation factor are
derived. Using these necessary conditions and the algorithms for the
determination of the inflation factor, an iterative, numerical
algorithm for the joint optimization is proposed. Some interesting
observations are made from the numerical results obtained from the
algorithm. Furthermore, the high-SNR sum-rate scaling factor
achievable over the CRC with imperfect CSIT is obtained.
\end{abstract}

% do the keywords

% start the main text ...
%----------------------------------------------------------------------
% SECTION I: Introduction
%----------------------------------------------------------------------
\section{Introduction}
\PARstart The cognitive radio channel (CRC) was introduced in
\cite{Devroye}. A cognitive radio is a device that can sense its
environment in real time and can accordingly adapt its transmission
strategy. These are of current interest because of the dramatically
high spectral efficiency they can achieve \cite{Devroye}. In
\cite{Devroye}, the authors introduced a more general cognitive
protocol under which the CRC is an interference channel with
degraded message sets \cite{C1}.

The Gaussian multiple-input multiple-output (MIMO) interference
channel consists of two transmitter-receiver pairs with each
transmitter having a message for its paired receiver and the
received signals are defined via equations $Y_1 = H_{11} X_1 +
H_{21} X_2 + Z_1$, $Y_2 = H_{12} X_1 + H_{22} X_2 + Z_2$. Here,
$\{H_{ij}\}_{i,j=1}^2$ are the fading channel matrices of dimensions
$r_j \times t_i$; the transmitted signals $X_1 \sim
\mathcal{C}\mathcal{N}(0,\Sigma_1)$ and $X_2 \sim \mathcal{C}
\mathcal{N}(0,\Sigma_2)$ are subject to the power
constraints\footnote{Notation: For a square matrix $A$,
$\mathrm{tr}(A)$, $|A|$, and $\mathrm{rank}(A)$ denote its trace,
determinant, and rank, respectively. For any general matrix $A$,
$A^*$ and $A^+$ denote the complex-conjugate transpose and
pseudo-inverse of matrix $A$, respectively. $\mathrm{vec}(A)$
denotes the vector obtained by stacking the columns of $A$. $I_m$ is
the $m \times m$ identity matrix. $\mathbb{E}_H$ denotes expectation
over the random variable $H$.} of $\mathrm{tr}(\Sigma_1) \leq P_1$
and $\mathrm{tr}(\Sigma_2) \leq P_2$; and $Z_1 \sim
\mathcal{C}\mathcal{N}(0,I_{r_1})$ and $Z_2 \sim
\mathcal{C}\mathcal{N}(0,I_{r_2})$ are the additive noises
\cite{C1}. The Gaussian MIMO CRC is defined as the Gaussian MIMO
interference channel in which the second transmitter (corresponding
to signal $X_2$) or the cognitive transmitter (CT) knows the message
(or the codeword) of the first or the primary transmitter
(corresponding to signal $X_1$) non-causally \cite{Devroye}.

An achievable-rate region for the Gaussian MIMO CRC has been
proposed in \cite{Sridharan}. In their coding scheme, the CT,
because of its non-causal knowledge, acts as a relay to aid the
primary receiver and also transmits its own message to its paired
(or cognitive) receiver. It employs dirty paper coding (DPC)
\cite{Costa} to cancel the interference at the cognitive receiver
due to the signals intended for the primary receiver. With the
assumption that the required channel matrices are known at the
transmitters and the receivers, it is further shown in
\cite{Sridharan} that their achievable-rate region includes points
corresponding to the sum-capacity of the MIMO CRC under certain
conditions. Now, to achieve the sum-capacity, an optimization over
the transmit covariances is required. This problem is studied in
\cite{SriramRajiv} where the authors propose the so-called adaptive
sum-power iterative waterfilling algorithm which computes the
sum-capacity and the optimal transmit covariances.

Although the MIMO CRC is increasingly being studied under the
assumption of perfect transmitter-channel-knowledge (CSIT)
\cite{Jovicic}, \cite{Sridharan}, \cite{SriramRajiv},
\cite{SriramVerdu}, etc., not many papers \cite{Lin2} exist which
deal with the practically important scenario of imperfect-CSIT CRC.
We find it timely to consider the aforementioned problem of
covariance optimization under imperfect CSIT. Towards this end, one
needs to seek answers to the following two questions:

1) \emph{DPC at the CT under imperfect CSIT: } Since the channel
seen by the cognitive receiver is of the form $Y=H_1 X+H_2 S +Z$
(this will become more clear in Section \ref{CovOpt}), where $S$ is
the interference known non-causally at the CT but not at its
receiver, it is imperative to first study the problem of DPC over
this channel when there is imperfect CSIT of $H_1$ and $H_2$. This
problem is equivalent to the determination of the optimal inflation
factor (see \cite{Costa}) under imperfect CSIT. We studied a similar
problem for the fading dirty paper channel (FDPC) $Y=H(X+S)+Z$ in
\cite{Vaze} and developed two iterative algorithms for determination
of inflation factor. These algorithms significantly improve the
prior attempts mentioned therein. The same problem for the channel
$Y=H_1 X+H_2 S + Z$, which we call the Generalized FDPC (G-FDPC),
has been considered in \cite{Mitran}, but only in the special case
of (all) single-antenna terminals, and a suboptimal solution is
proposed. We study this important problem in Section \ref{G-FDPC} of
this paper.

2) \emph{Covariance optimization under imperfect CSIT: } The problem
of covariance optimization is considerably involved, even under
perfect CSIT. In \cite{SriramRajiv}, rather than using the
achievable-rate region of \cite{Sridharan}, the authors formulate
the problem in terms of an outer-bound (obtained in
\cite{Sridharan}) to the capacity region which includes points
corresponding to the sum-capacity of the MIMO CRC. This is a
non-convex optimization problem and is converted into an equivalent
convex-concave game using the `MAC-BC' transformations
\cite{Sriram}. Then, for the resulting optimization, an iterative
numerical algorithm is proposed. Unfortunately, the algorithm can
not always guarantee the optimum solution.

The use of the outer-bound or the `MAC-BC' transformations is not
possible under imperfect CSIT. Also, unlike the perfect-CSIT case
(under which the interference can be assumed to be canceled
perfectly by DPC), under imperfect CSIT, an additional optimization
over the inflation factor needs to be performed \emph{jointly} with
the transmit covariances. Furthermore, the problem becomes more
complicated because the sum-rate optimal solution need not
necessarily have the power constraints satisfied with equality (this
point is detailed later). Thus the imperfect-CSIT version of this
problem is also quite challenging.

A slightly different (due to a constraint on the rate of primary
user) version of the problem is considered in \cite{Lin2},
\cite{Lin3} for the CRC with all single-antenna terminals. The
authors of \cite{Lin2}, \cite{Lin3} consider the amplify-and-forward
strategy for relaying at the CT, and in this sense, their scheme is
less general than the one studied here. %

\section{DPC over the G-FDPC} \label{G-FDPC}
Motivation to study this problem will become more clear in Section
\ref{CovOpt}. But, as noted before, this is an important step in the
overall joint optimization. The G-FDPC is defined via equation
$Y=H_1 X + H_2 S + Z$. Here, $H_1$ and $H_2$ are the channel
matrices of dimensions $r \times t_x$ and $r \times t_s$,
respectively; the transmitted signal $X \sim \mathcal{C}\mathcal{N}
(0,\Sigma_X)$ has a power constraint of $P$; the interference $S
\sim \mathcal{C}\mathcal{N} (0,\Sigma_S)$ is known non-causally at
the transmitter but not at the receiver; $Z \sim
\mathcal{C}\mathcal{N} (0,\Sigma_Z)$ is the additive noise; and $X$,
$S$, and $Z$ are independent. Assume perfect receiver channel
knowledge but imperfect CSIT \footnote{We assume that the
transmitter only knows the distribution $H$. The case of partial
CSIT can be handled similarly.}. Assume $|\Sigma_X|$,
$|\Sigma_Z|>0$; let $\mathrm{tr}(\Sigma_S)=Q$,
$\mathrm{tr}(\Sigma_Z)=N$. Define $\mathrm{SNR}=\frac{P}{N}$
\footnote{Note that $N$ is total noise power.}. Select the auxiliary
random variable (see \cite{G-P} for definition) as $U=X+WS$, i.e.,
Costa's choice \cite{Costa} extended to the MIMO case, where the
$t_x \times t_s$ matrix $W$ is the inflation factor \footnote{Matrix
$W$ is called the inflation factor so as to be consistent with the
terminology introduced by Costa \cite{Costa}.}. Similar to
\cite{Vaze}, we obtain the achievable rate as given by
\begin{eqnarray}
R = \max_W \mathbb{E}_H \log \frac{|\Sigma_X| |\Sigma_Z + H_1
\Sigma_X H_1^* + H_2 \Sigma_S H_2^*|}{| \mathrm{M} |} \label{Rach}
\end{eqnarray}
with $\hspace{-1pt}\mathrm{M} \hspace{-2pt}= \hspace{-3pt}
\left[\hspace{-6pt}
\begin{array}{cc}
    \Sigma_X + W \Sigma_S W^* \hspace{-6pt} & \Sigma_X H_1^* + W \Sigma_S H_2^* \\
    H_1 \Sigma_X + H_2 \Sigma_S W^* \hspace{-6pt} & \Sigma_Z
+ H_1 \Sigma_X H_1^* + H_2 \Sigma_S H_2^* \\
    \end{array} \hspace{-6pt} \right]$, and $H=[H_1 \hspace{2pt}
    H_2]$. The above rate expression is valid only if $|\Sigma_X|>0$. The case
of $|\Sigma_X|=0$ can be handled as in \cite{Vaze2}. We define the
no-interference upper-bound $R_{\mathrm{noS}}$ as the rate
achievable over the G-FDPC in absence of interference (i.e., when
$Q=0$) or $R_{\mathrm{noS}}=\mathbb{E}_{H_1} \log
\frac{|\Sigma_Z+H_1 \Sigma_X H_1^*|}{|\Sigma_Z|}$.

\begin{figure} \hspace{-12pt}
\includegraphics[height=2.83in,width=3.7in]{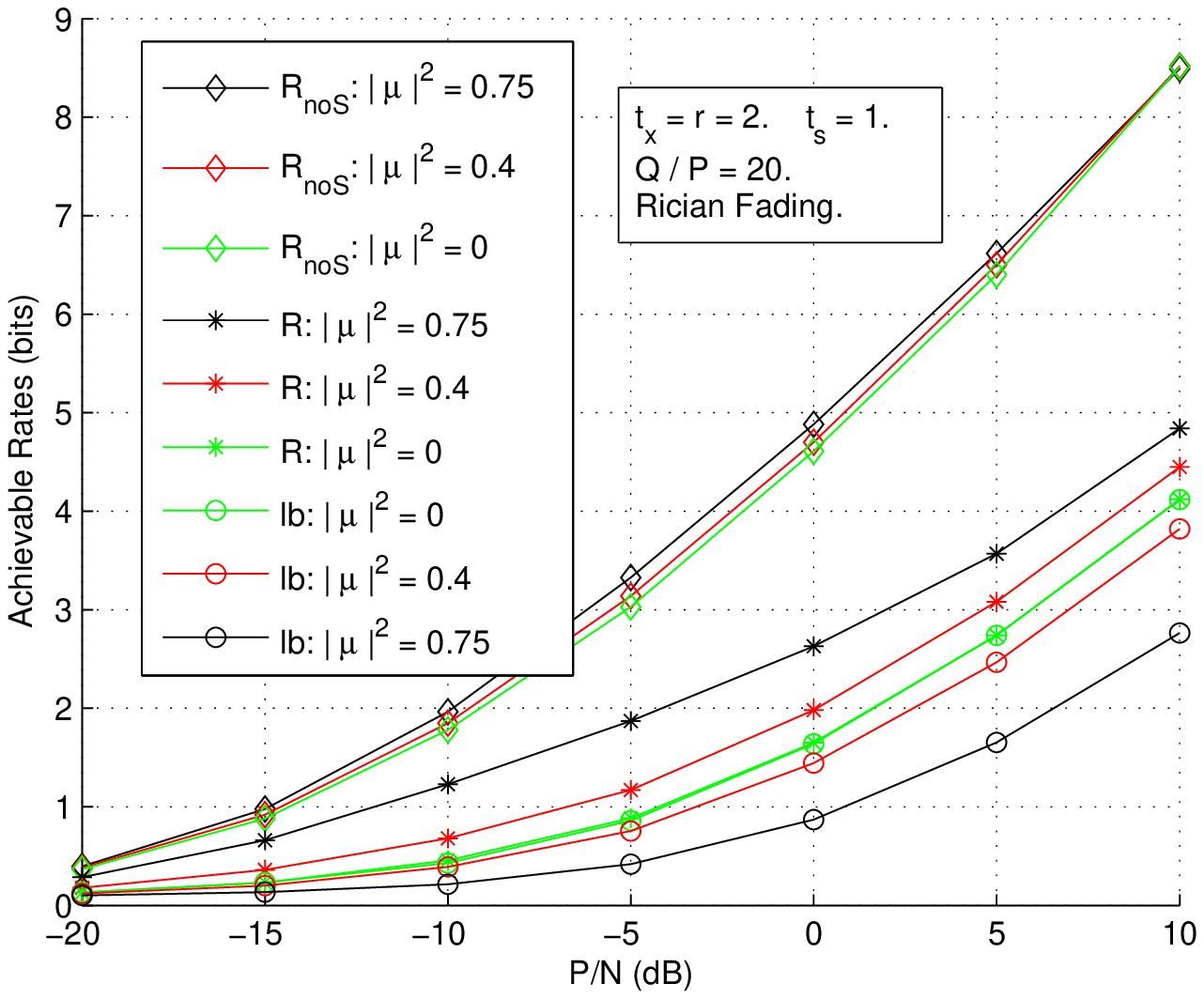}
\caption{Achievable Rates vs. SNR: Rician Fading.} \label{GFDPC1}
\end{figure}

The problem of determination of inflation factor, i.e., the
maximization in (\ref{Rach}) is equivalent to $\min_W
\mathbb{E}_{[H_1 \hspace{2pt} H_2]}\log |\mathrm{M}|$. As noted in
\cite{Vaze}, this is a non-convex optimization problem, and it seems
intractable to obtain a closed-form solution. It is possible however
to generalize our algorithms in \cite{Vaze} developed for the FDPC
to the G-FDPC. Due to lack of space, we discuss here the basic idea
and omit the details.

\begin{figure*}[!b]
\begin{picture}(10,1)
\put(10,10){\line(1,0){500}}
\end{picture}
\begin{eqnarray}
R_{sum} = \mathbb{E}_{\bar{H}} \bigg\{ \hspace{-1pt} \log
\frac{|I_{r_1} + \bar{H}_1 T_1 T_1^* \bar{H}_1^* + H_{21} T_2 T_2^*
H_{21}^* |} {| I_{r_1} + H_{21} T_2 T_2^* H_{21}^* | } \hspace{-1pt}
+ \hspace{-1pt} \log \frac{ | I_{r_2} + H_{22} T_2 T_2^* H_{22}^* +
\bar{H}_2 T_1 T_1^* \bar{H}_2^*|} {\left| \hspace{-4pt}
\begin{array}{cc}
I_{t_2} +W T_1 T_1^* W^* \hspace{-2pt} & T_2^* H_{22}^* + W T_1 T_1^* \bar{H}_2^*\\
H_{22} T_2 + \bar{H}_2 T_1 T_1^* W^* \hspace{-2pt} & I_{r_2} + H_{22} T_2 T_2^* H_{22}^* +  \bar{H}_2 T_1 T_1^* \bar{H}_2^* \\
    \end{array} \hspace{-4pt}
    \right| } \bigg\} \label{sum-rate_T1_T2}
\end{eqnarray}
\begin{picture}(10,1)
\put(10,10){\line(1,0){500}}
\end{picture}
\begin{eqnarray}
\lefteqn{ \left[
  \begin{array}{cc}
    \lambda_p^{-1} I_{t_1} & 0 \\
    0 & \lambda_c^{-1} I_{t_2}  \\
  \end{array}
\right] T_1 = \mathbb{E}_{\bar{H}} \left\{ \bar{H}_1^* N_1^{-1}
\bar{H}_1^* + \bar{H}_2^* N_2^{-1} \bar{H}_2^* - [ W^* \hspace{3pt}
\bar{H}_2^*] D_2^{-1} \left[
\begin{array}{c}
    W \\
    \bar{H}_2 \\
    \end{array}
\right] \right\} T_1 = g_1(T_1, \hspace{1pt} T_2, \hspace{1pt} W) , }\label{NecCondns1} \\
&& {} \hspace{-4pt} \lambda_c^{-1} T_2 = \mathbb{E}_{\bar{H}}
\left\{ H_{21}^* N_1^{-1} H_{21} T_2 - H_{21}^* D_1^{-1} H_{21} T_2
+ H_{22}^* N_2^{-1} H_{22} T_2 - [0 \hspace{4pt} H_{22}^*] D_2^{-1}
[I_{t_2} \hspace{4pt} T_2^* H_{22}^*]^* \right\}
\label{NecCondns2} \\
&& {} \hspace{10pt} = g_2(T_1, \hspace{1pt} T_2, \hspace{1pt} W),
\hspace{45pt} \cdots \mbox{ where } N_1 = I_{r_1} + \bar{H}_1 T_1
T_1^* \bar{H}_1 + H_{21} T_2 T_2^* H_{21}^*, \hspace{2pt} D_1 =
I_{r_1} + H_{21} T_2
T_2^* H_{21}^*, \nonumber \\
&& {} \hspace{-4pt} N_2 = I_{r_2} + H_{22} T_2 T_2^* H_{22}^* +
\bar{H}_2 T_1 T_1^* \bar{H}_2^*, \mbox{ and $D_2$ is the
block-partitioned matrix in equation (\ref{sum-rate_T1_T2}).}
\nonumber
\end{eqnarray}
\end{figure*}

In the first algorithm, we minimize the objective function stepwise,
i.e., at each step, we minimize over only one row of $W$, while
treating all other rows as constants. Note that only the $k^{th}$
row and the $k^{th}$ column of matrix $M$ depend on the $k^{th}$ row
of $W$. Therefore, the minimization over one row of $W$ (while
treating other rows as constants) can be done analytically if the
objective function is upper-bounded by moving the expectation inside
the logarithm. Thus, one iteration of the algorithm consists of
successive (stepwise) minimizations over all rows of $W$, and these
iterations are repeated until a good choice is obtained.

In the second algorithm, we solve for the stationary point of the
objective function, i.e., solve an equation $\frac{d}{dW}
\mathbb{E}_{[H_1 \hspace{2pt} H_2]} \log |\mathrm{M}| = 0$. Using
the obtained necessary conditions, an iterative algorithm is
proposed.

\begin{figure} \hspace{5pt}
\includegraphics[height=2.25in,width=3.2in]{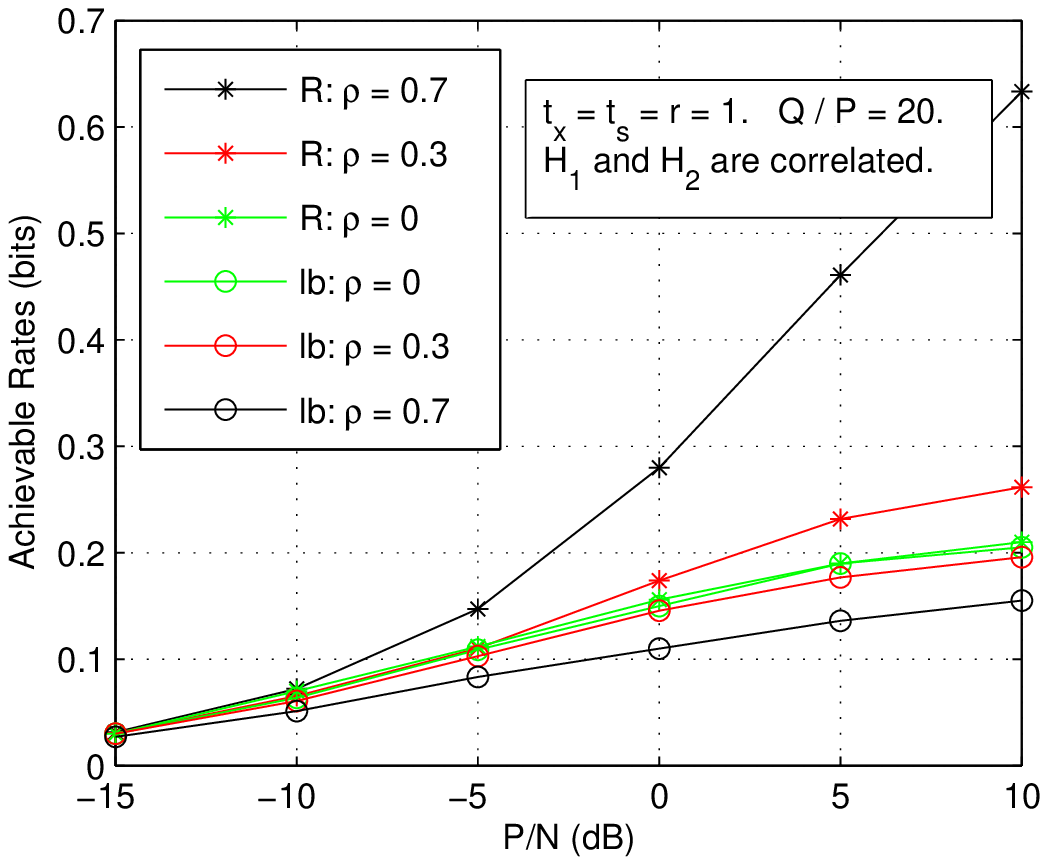}
\caption{Achievable Rates vs. SNR: $H_1$ and $H_2$ Correlated.}
\label{GFDPC2}
\end{figure}

\emph{\underline{Numerical Results:} } Here, `$\mathrm{lb}$' denotes
the rate achievable using $W=0$, i.e., by treating the interference
as noise. $R$ denotes the rate achievable using the algorithms. In
Fig. \ref{GFDPC1}, we take $H_1$ and $H_2$ to be independent with
their elements $\sim$ i.i.d. $\mathcal{C} \mathcal{N}(\mu,
\sigma^2)$ with $|\mu|^2+\sigma^2=1$ and $\mu =
|\mu|\frac{(1+j)}{\sqrt{2}}$. When $\mu=0$, the all-zero inflation
factor performs almost as well as the inflation factor obtained
using the algorithms. However, as $|\mu|$ increases, the algorithms
outperform the simple choice of $W=0$. This type of observation was
also made in \cite{Mitran} in the case of SISO G-FDPC. It is
generalized here to the MIMO case. In Fig. \ref{GFDPC2}, the fading
coefficients are correlated, i.e., $H_1$, $H_2 \sim
\mathcal{C}\mathcal{N}(0,1)$ with $\rho=E(H_1^*H_2)$. As $\rho$
increases from $0$ to $0.7$, the algorithms perform better than
simply setting $W=0$.

A considerable difference between $R$ and $R_{\mathrm{noS}}$ is seen
in Fig. \ref{GFDPC1}. It should be noted that $R_{\mathrm{noS}}$
corresponds to the perfect interference-cancelation, while the curve
$R$ is for no CSIT. The gap between the two can be bridged with the
availability of partial CSIT. Additionally, $R_{\mathrm{noS}}$ is
loose in the high-SNR regime because of the difference in the
achievable scaling factors of $R$ and $R_{\mathrm{noS}}$ (see
Theorem 1).

Loosely speaking, it appears that for DPC to perform significantly
better than the naive scheme of treating the interference as noise,
it is necessary to have the matrix $E(\mathrm{vec}(H_1)^*
\mathrm{vec}(H_2))$ `non-zero', i.e., to have $H_1$ and $H_2$
`correlated'. The `more' non-zero the above matrix is (or the `more
highly' $H_1$ and $H_2$ are correlated), the greater is the
improvement. We believe this to be the fundamental nature of DPC
over the G-FDPC under imperfect CSIT. Also see the discussion
following Theorem 1.

\section{Optimization over the Transmit Covariances} \label{CovOpt}
As per the coding scheme of \cite{Sridharan}, let
$X_2=X_{21}+X_{22}$ where the signal $X_{21}$ corresponds to
relaying and is correlated with $X_1$ while $X_{22}$ is the signal
intended for the cognitive receiver. Let $\left[ \hspace{-4pt}
\begin{array}{c}
    X_1 \\
    X_{21} \\
\end{array} \hspace{-4pt}
\right] \sim \mathcal{C}\mathcal{N}\left( 0,\Sigma = \left[
\hspace{-4pt} \begin{array}{cc}
    \Sigma_1 & V \\
    V^* & \Sigma_{21} \\
    \end{array} \hspace{-4pt} \right] \right)$, and $X_{22} \sim
\mathcal{C}\mathcal{N}(0,\Sigma_{22})$. Also let $\Sigma=T_1T_1^*$
and $\Sigma_{22}=T_2 T_2^*$ for some $T_1$ and $T_2$; and
$X_{22}=T_2 X_{22}'$ with $X_{22}' \sim
\mathcal{C}\mathcal{N}(0,I_{t_2})$. The CT would choose the
auxiliary random variable as $U=X_{22}' + W \left[
\begin{array}{c}
X_1 \\
X_{21} \\
\end{array}
\right]$, where $X_{22}'$ is independent of $X_1$ and $X_{21}$.
Hence $\Sigma_2=\Sigma_{22} + \Sigma_{21}$.

Now the channel between the CT-receiver pair is $Y_2 = H_{22} X_{22}
+ [H_{12} \hspace{2pt} H_{22}] \left[ \hspace{-4pt}
                       \begin{array}{c}
                         X_1 \\
                         X_{21} \\
                       \end{array}
 \hspace{-4pt}   \right] + Z_2$ which resembles the G-FDPC.
Therefore, using the algorithms of Section \ref{G-FDPC}, we can
determine the inflation factor to be used at the CT once $\Sigma$
(or $T_1$) and $\Sigma_{22}$ (or $T_2$) are specified. This explains
the reason to first study DPC over the G-FDPC.

Denote $\bar{H}_1 =[H_{11} \hspace{4pt} H_{21}]$, $\bar{H}_2 =
[H_{12} \hspace{4pt} H_{22}]$, and $\bar{H} = [\bar{H}_1^*
\hspace{2pt} \bar{H}_2^*]^*$. Then the achievable sum-rate
$R_{sum}=R_p+R_c$ under no CSIT and perfect receiver channel
knowledge is given by equation (\ref{sum-rate_T1_T2}) at the bottom
of the page.

Since $W$ depends on $T_1$ and $T_2$, we need to optimize $R_{sum}$
jointly over $T_1$, $T_2$, and $W$, as mentioned earlier. However,
since $W$ can be determined given $T_1$ and $T_2$, let us first
consider the optimization of $R_{sum}$ over $T_1$ and $T_2$ for a
given value of $W$; later the algorithm for the joint optimization
can be formulated. Let us consider: $\max_{T_1,T_2} R_{sum}$,
subject to $\mathrm{tr}(\Sigma_1) \leq P_p$ and
$\mathrm{tr}(\Sigma_{21}+\Sigma_{22}) \leq P_c$. This is a
non-convex optimization problem. To obtain the necessary conditions,
we form the lagrangian $J$; and set $\frac{\partial J}{\partial T_1}
= 0$ and $\frac{\partial J}{\partial T_2} = 0$. We omit the details
of differentiation and directly state the necessary conditions,  as
given by equations (\ref{NecCondns1}) and (\ref{NecCondns2}) at the
bottom of the page, where $\lambda_p^{-1}$ and $\lambda_c^{-1}$ are
the lagrange multipliers.

\emph{\underline{Algorithm for the Joint Optimization (Alg. 1):} }
\begin{enumerate}
\item Start with some initial choices $T_1^{(0)}$ and $T_2^{(0)}$.
For these choices, determine $W^{(0)}$ using the algorithms
discussed in Section \ref{G-FDPC}.
\item At the $n^{th}$ iteration,
\begin{itemize}
  \item Determine the transmit covariances: to this end, we set $T_1^{(n)} = \left[
  \begin{array}{cc} \hspace{-5pt}
    \lambda_p I_{t_1} \hspace{-5pt} & 0 \\
    0 \hspace{-5pt} & \lambda_c I_{t_2} \\
\end{array} \hspace{-5pt}
\right] g_1(T_1^{(n-1)}, T_2^{(n-1)}, W^{(n-1)})$ and $T_2^{(n)} =
\lambda_c g_2(T_1^{(n-1)}, T_2^{(n-1)}, W^{(n-1)})$. The required
expectations are evaluated numerically. Find lagrange multipliers so
as to meet the power constraints.
\item For $T_1^{(n)}$ and $T_2^{(n)}$ obtained above,
determine $W^{(n)}$.
\end{itemize}
\item Repeat the above step until the increase in the achievable
sum-rate is negligible.
\end{enumerate}

The statement above regarding the determination of lagrange
multipliers warrants a discussion. Note, the power transmitted by
either transmitter increases with $\lambda_{p,c}$. Therefore, the
feasible region for the $\lambda$'s is of the form $0 <
\lambda_{p,c} \leq \lambda_{p,c} ^{max}$, where
$(\lambda^{max}_p,\lambda_c^{max})$ is a point at which both the
power constraints are satisfied with equality. One can expect the
optimal point to be $(\lambda^{max}_p,\lambda_c^{max})$ at which
both the transmitters operate with the maximum available power.
However, since the signal intended for the primary receiver is an
interference for the cognitive receiver and vice versa, the sum-rate
need not necessarily be a nondecreasing function of either
$\lambda_p$ or $\lambda_c$. Hence, the optimal point for
$\lambda$'s, i.e., $\lambda_{p,c}^{opt}$ can be any interior or
boundary point of the above rectangular region. Note, the choice of
$\lambda$'s dictates the covariance matrices, and therefore the
inflation factor. Considering the fact that only an algorithmic
solution is available for the inflation factor and all the required
expectations need to be evaluated numerically, the problem of
determination of optimal $\lambda$'s looks intractable. In the
numerical examples, we consider a suboptimal solution of solving the
power constraints as strict equalities.

We have developed one more algorithm (Alg. 2) for the joint
optimization which serves as a lower-bound on the rate achievable
using Alg. 1.
\begin{enumerate}
\item Assume that $T_1=0$. Determine $T_2$ to maximize $R_c$ subject to
$\mathrm{tr}(\Sigma_{22}) = \frac{P_c}{2}$ (note the equality here).
\item For given $T_2$, determine $T_1$ to maximize $R_p$ under the
constraints that $\mathrm{tr}(\Sigma_1) = P_p$ and
$\mathrm{tr}(\Sigma_{21}) = \frac{P_c}{2}$.
\item For given $T_1$, determine $T_2$ and $W$ to maximize $R_c$
under the constraint that $\mathrm{tr}(\Sigma_{22}) =
\frac{P_c}{2}$.
\item Repeat Steps 2 and 3 above until the increase in the
achievable rate is negligible.
\end{enumerate}
Thus $R_p$ and $R_c$ are maximized here greedily over $T_1$ and
$(T_2$, $W)$, respectively. For these maximizations, the algorithm
of joint optimization developed in \cite{Vaze2} is used.

\section{High-SNR Analysis: Scaling Factor} \label{ScalingFactor}
\newtheorem{theorem}{Theorem}
\begin{theorem}{\emph{\underline{G-FDPC}:}}
Assume that the ratio $\frac{Q}{P}$ is constant as $P \to \infty$,
and the fading processes are such that for any positive
semi-definite $A$, $\mathrm{rank}([H_1 \hspace{2pt} H_2] A [H_1
\hspace{2pt} H_2]^*) = \min (r, \mathrm{rank}(A))$ with probability
$1$. The high-SNR scaling factor achievable over the no-CSIT G-FDPC
using DPC is independent of the choice of $W$, as long as $W$ is
chosen such that the term $\log |\Sigma_X + W\Sigma_S W^*|$ scales
in the high-SNR regime as $t_x \log$ SNR.
\end{theorem}

Thus, the naive scheme of treating the interference as noise (i.e.,
$W=0$) achieves the optimal scaling factor, which is given by
$\min(r,\mathrm{rank}(\Sigma_X) + \mathrm{rank}(\Sigma_S)) -
\min(r,\mathrm{rank}(\Sigma_S))$. Also note that there is no loss of
generality in choosing $W$ to satisfy the condition stated in
Theorem 1 because $W$'s that do not satisfy this condition can
achieve only a suboptimal scaling factor.

The intuition detailed in the paragraph just preceding Section
\ref{CovOpt} can explain the result of Theorem 1. Consider the FDPC
with $|\Sigma_X| >0$, or equivalently, the G-FDPC with $H_1=H_2=H$
and $t_x=t_s=t$. The high-SNR scaling factor of $\min(t,r)$, which
is equal to that of the corresponding no-interference upper-bound,
is achievable with the choice of $W=I_t$ \cite{Vaze}. Next consider
the FDPC with $|\Sigma_X|=0$ (let $\Sigma_X =TT^*$); this channel is
then equivalent to the G-FDPC $Y=H_1X' + H_2 S + Z$ with $H_1 = HT$,
$H_2=H$, and $X' \sim \mathcal{C}\mathcal{N}(0,I)$ (so $H_1$ and
$H_2$ not equal). In this case, the achievable scaling factor may
not always be equal to that of the no-interference upper-bound; but
in most cases, by making an appropriate choice for $W$ (say, $W=T^+$
\cite{Vaze2}), one can achieve a better scaling factor than that
achievable with $W=0$. Note, the G-FDPCs in the two cases above do
not satisfy the assumption regarding $H_1$ and $H_2$ made in Theorem
1. Finally, consider the G-FDPC that satisfies the assumption of
Theorem 1 (for example, $H_1$ and $H_2$ are independent and
Rayleigh-faded). Then, as per Theorem 1, there is no advantage in
optimizing over $W$ as far as the scaling factor is concerned. Thus,
the `more highly' $H_1$ and $H_2$ are correlated, the `larger' is
the increase in the scaling factor over that achievable by treating
the interference as noise.

\begin{theorem}{\emph{\underline{CRC}:}}
Assume that the channel matrices $\{H_{ij}\}$ are full rank and
independent; and the ratio $\frac{P_p}{P_c}$ remains constant. The
high-SNR($=P_p$) sum-rate scaling factor achievable over the no-CSIT
CRC is given by \vspace{-1pt}
\begin{eqnarray*}
\lefteqn{ \gamma_{sum} = \max_{\mathrm{rank}(\Sigma'), \hspace{1pt}
\mathrm{rank}(\Sigma_{22})} \gamma_p + \gamma_c \mbox{ with} } \\
&& {} \gamma_p = \min (r_1,\mathrm{rank}(\Sigma')) - \min (r_1,
\mathrm{rank}(\Sigma_{22})), \\
&& {} \gamma_c = \min(r_2, \mathrm{rank}(\Sigma')) -
\min(r_2,\mathrm{rank}(\Sigma')- \mathrm{rank}(\Sigma_{22})),
\end{eqnarray*}
where $\Sigma'$ is the covariance matrix of $\left[ \begin{array}{c}
      X_1 \\
      X_2
    \end{array} \right]$; and the maximization is under the
constraints of $ 0 < \mathrm{rank}(\Sigma') \leq t_1+t_2$, and $0
\leq \mathrm{rank}(\Sigma_{22}) \leq \min (t_2,
\mathrm{rank}(\Sigma'))$.
\end{theorem}
\begin{proof}
We present the outline here and omit the details. Given $\Sigma$ and
$\Sigma_{22}$, $R_p$ achieves the scaling factor of $\gamma_p$
because the primary receiver treats the interference $X_{22}$ as
noise. $\gamma_c$ is achieved by the choice of $W = [0 \hspace{4pt}
T_2^+]$ (see Theorem 1 above and Theorem 1 of \cite{Vaze2}).
\end{proof}
The maximization in Theorem 2 is over only finitely many values;
thus, can be done via exhaustive search. Note, the power constraints
are not solved as strict equalities here.

\section{Numerical Results for the CRC}
In figures, `ub' denotes the sum-rate achievable by optimizing over
the transmit covariances under the assumption that the interference
is perfectly canceled at the cognitive receiver (i.e., the sum-rate
with $R_c = E_{H_{22}} \log |I_{r_2} + H_{22} \Sigma_{22}
H_{22}^*|$). We quantize each element of the fading matrices
separately using an `equally spaced level' quantizer as defined in
\cite{Max}. In figures, if $B = [B_{ij}]$, then $B_{ij}$ denotes the
number of feedback bits used per element of matrix $H_{ij}$.
Further, $H_{ij}$ are independent. Alg. 1 is unfortunately sensitive
to the initial choices. We take $4$ to $5$ initial choices in these
examples and then select the best solution.

\begin{figure} \hspace{-8pt}
\includegraphics[height=2.5in,width=3.54in]{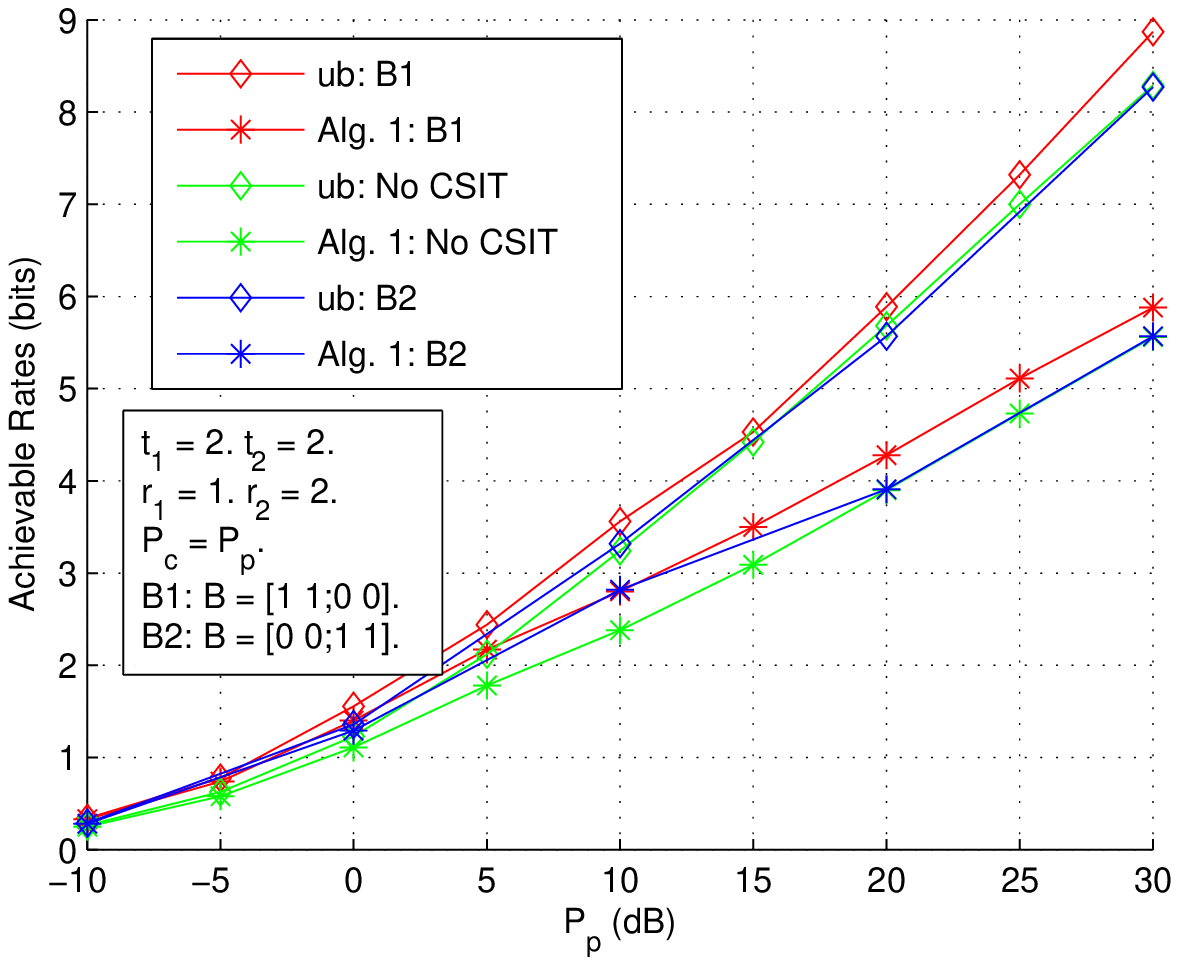}
\caption{Achievable sum-rate vs. Pp.} \label{CRC1}
\end{figure}

\begin{figure} \hspace{10pt}
\includegraphics[height=2.1in,width=3in]{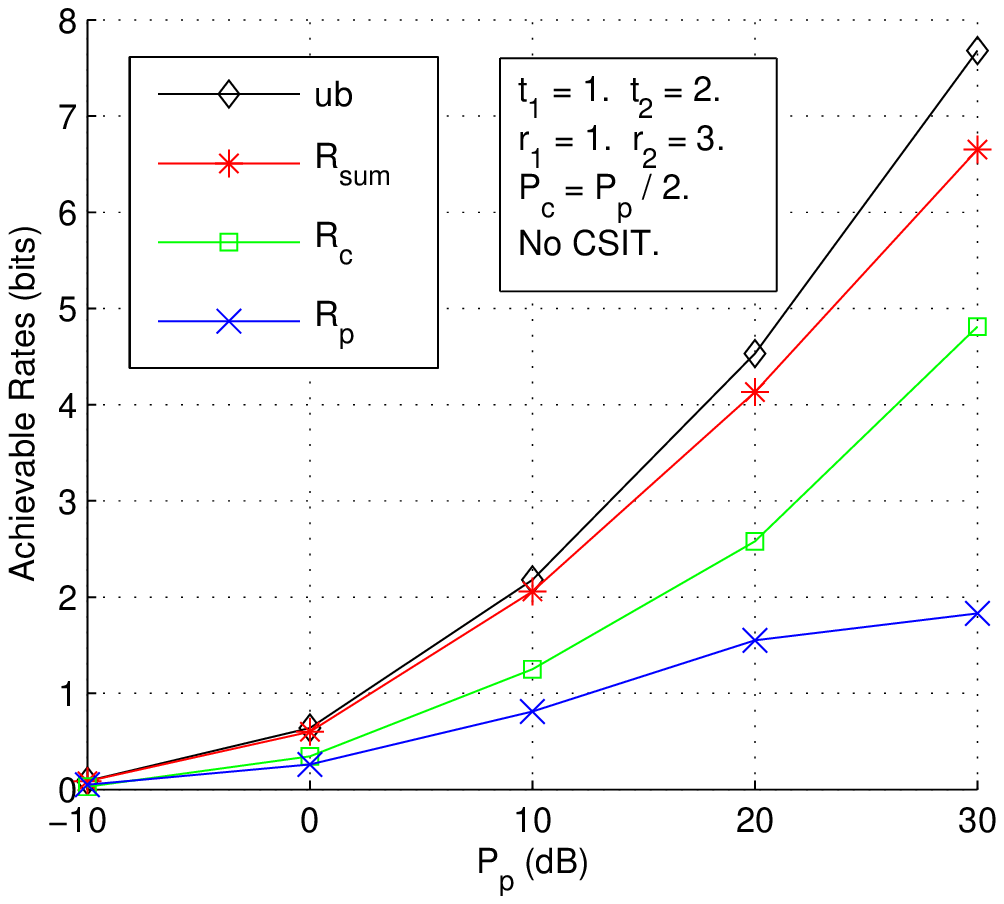}
\caption{Achievable sum-rate vs. Pp.} \label{CRC2}
\end{figure}

\begin{figure}
\includegraphics[height=2.2in,width=3.1in]{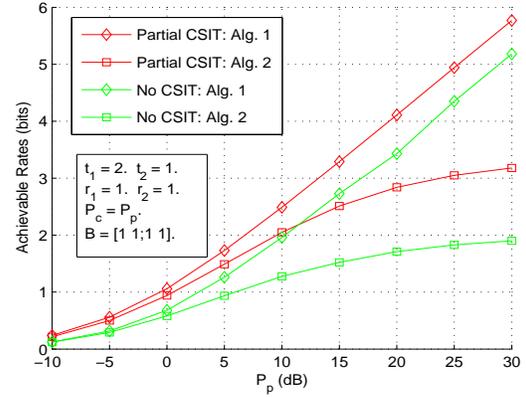}
\caption{Comparison of Alg. 1 and Alg. 2.} \label{CRC3}
\end{figure}

In Fig. \ref{CRC1}, we consider the CRC with elements of $\{H_{ij}\}
\sim$ i.i.d. $\mathcal{N}(0.6,0.64)$. The improvement in the
achievable sum-rate with the introduction of partial CSIT is
evident. Here, the scaling factor of $1$ is achieved for $R_{sum}$
by letting the CT to use its entire power for relaying. Hence, the
curve corresponding to B2 merges with that corresponding to the no
CSIT at high SNR. For the CRC of Fig. \ref{CRC2}, we have the
elements of $\{H_{ij}\} \sim$ i.i.d. $\mathrm{Unif}[0,1]$. For this
CRC, as per Theorem 2, the optimal solution should achieve
$\gamma_{sum}=2$ with $\gamma_p=0$. This fact can be easily seen
from the plot. In Fig. \ref{CRC3}, we have the CRC with elements of
$\{H_{ij}\} \sim$ i.i.d. $\mathcal{N}(0,1)$. It can be seen that
Alg. 1 outperforms Alg. 2. However, in some cases, for example, the
CRCs in Figs. \ref{CRC1} and \ref{CRC2}, Alg. 2 does provide a
relatively tight lower-bound. Coming back to Fig. \ref{CRC1} again,
$R_{sum}$ achieves the scaling factor of $1$ whereas according
Theorem 2, $\gamma_{sum}=2$. This is achieved by setting $\Sigma_1 =
0$, i.e., the primary transmitter needs to turn off its power. The
apparent inconsistency here is because we have considered a
suboptimal solution of solving the power constraints as strict
equalities. This example emphasizes the importance of the problem of
determination of $\lambda's$.

\section{Conclusion}
This paper is one of the earliest works that studies the
imperfect-CSIT MIMO CRC. To the best of the authors' knowledge, it
proposes for the first time a transmission strategy for the
multi-antenna CRC with imperfect CSIT. En-route, brings into focus
the problem of determination of $\lambda$'s. Furthermore, the paper
derives an achievable high-SNR sum-rate scaling factor. It would be
worthwhile to obtain the highest-achievable sum-rate scaling factor.
This problem can be interesting; recall its counterpart for the
Gaussian MIMO broadcast channel, a problem that is open even after
serious attempts. More efforts are needed to answer these two open
questions.

% do the biliography:
\bibliographystyle{IEEEbib}
\bibliography{CRC_finalversion}

\end{document}